\documentclass[12pt]{iopart}

\usepackage[english]{babel}
\usepackage{amssymb}
\usepackage{epsfig}
\def\lsim{\raisebox{-.4ex}{$\stackrel{<}{\scriptstyle \sim}$\,}}

\begin{document}
\newcommand{\A}{{\mathcal{A}}}
\newcommand{\dA}{\delta{\mathcal{A}}}
\newcommand{\Od}{{\cal O}}
\newcommand{\degree}{^\circ}
\newcommand{\K}{\textrm{K}}


\title{Viability of vector-tensor theories of gravity}%

\author{Jose Beltr\'an Jim\'enez and Antonio L. Maroto}

\address{Departamento de  F\'{\i}sica Te\'orica I,
Universidad Complutense de Madrid, 28040 Madrid, Spain.}
\date{\today}

\begin{abstract}
We present a detailed study of the viability of general vector-tensor
theories of gravity in the presence of an arbitrary temporal
background vector field. We find that there are six different
classes of theories which are indistinguishable from
General Relativity by means of local gravity experiments. We study the
propagation
speeds of scalar, vector and tensor perturbations and obtain
the conditions for classical stability of those models.
We compute the energy density of the different modes and
find the conditions for the absence of ghosts in the quantum theory.
We conclude that the only  theories  which can pass all
the viability conditions for arbitrary values of the 
background vector field are not only those of the pure Maxwell type,
but also  Maxwell theories supplemented with a (Lorentz type) gauge
fixing term.

\end{abstract}

\maketitle 

\section{Introduction}
Although the interest in alternative gravity theories has been
present over the years \cite{MTW,Will}, it has been recently
renewed by the unexpected observations made in cosmological and
astrophysical contexts. Indeed, the difficulties found by the
Standard Model (SM) of elementary particles and General Relativity
(GR) in order to explain the nature of dark matter and dark energy
suggest either the need for new physics beyond the SM or the
modification of GR at very large scales.

Among such modifications, we find the so called vector-tensor
theories of gravity, proposed in the early 70's \cite{VT1,VT2},
and where, in addition to the metric tensor, a dynamical vector
field is introduced in such a way that it only couples to matter
through gravitational interactions.

After some initial proposals, these theories were practically
forgotten due to consistency problems and only  
particular classes of models with fixed norm vector or potential terms 
 have been considered in the context of violations of Lorentz invariance (see
\cite{Jacobson} and references therein). However, very recently, it 
has been shown \cite{BM,BMEM} that unconstrained vector-tensor theories
without potential terms are excellent candidates for dark energy.

The main three difficulties found in this type of theories were
the following \cite{FosterJacobson,Elliot}:

a) Inconsistencies with local gravity tests:
in addition to modifications in the static parametrized
post-Newtonian (PPN) parameters ($\gamma$, $\beta$), these
theories typically predict preferred frame effects, parametrized
by $(\alpha_1,\alpha_2)$. Such effects are produced by the motion
with respect to the privileged frame selected by the presence of
the  background vector field. The increasing observational
precision in Earth and Solar System experiments
 seems to exclude this type of effects \cite{Will}.

b) Classical instabilities:
they are generated by the presence of modes with negative
propagation speed squared. This gradient instabilities give rise
to exponentially growing perturbations which make the theory
unphysical.

c) Quantum instabilities (ghosts):
they typically appear when the energy of some perturbation modes
becomes negative. In such a case it has been shown that the vacuum
state of the corresponding quantum theory is unstable
\cite{Cline}.

It has been also argued  \cite{CarrollLim} that the presence of
faster than light propagation modes could give rise to causality
inconsistencies although this conclusion is far from clear (see for
instance \cite{Mukhanov}) and
even some authors claim that consistency requires the presence of
such modes \cite{Elliot}. For that reason in this work we will not
include it as a viability requirement.

Although it is usually claimed that vector-tensor
theories are plagued by the mentioned consistency problems, the
actual situation is that none of them has been studied in detail,
and they are only based on qualitative arguments which, in many
cases, turn out to be inappropriate. In this work we
present a detailed analysis of the viability of unconstrained
vector-tensor theories and obtain the precise conditions
in order to satisfy the above requirements.

\section{Local gravity constraints}

Without any other restriction, apart from having second order linear 
equations of motion, the most general action for a vector-tensor
theory without potential terms  can be written as follows
\cite{Will}:
\begin{eqnarray}
S[g_{\mu\nu},A_\mu]=\int d^4x\sqrt{-g}\left[-\frac{1}{16\pi
G}R+\omega RA_\mu A^\mu\right.\nonumber\\+\left.\sigma
R_{\mu\nu}A^\mu A^\nu+\tau\nabla_\mu A_\nu \nabla^\mu
A^\nu+\varepsilon F_{\mu \nu}F^{\mu\nu}\right]\label{Action}
\end{eqnarray}
with $\omega$, $\sigma$, $\tau$, $\varepsilon$ dimensionless
parameters and $F_{\mu\nu}=\partial_\mu A_\nu-\partial_\nu A_\mu$.
Notice that the term including the Ricci tensor can be rewritten
as $\nabla_\mu A^\mu\nabla_\nu A^\nu-\nabla_\mu A^\nu\nabla_\nu
A^\mu$. These two terms do not appear independently in the 
most general action because they can be recasted in the terms
written in (\ref{Action}).

Since GR agrees with Solar System experiments with high precision,
it would be desirable that a viable vector-tensor theory had the
same set of PPN parameters as GR, i.e. $\gamma=\beta=1$ and
$\alpha_1=\alpha_2=0$. Notice that the rest of PPN parameters
vanishes identically for models described by (\ref{Action}). The
PPN parameters for general vector-tensor theories have been
calculated in \cite{Will} assuming the existence of a constant
background time-like vector field:
\begin{eqnarray}
 A_\mu=(A,0,0,0)\label{vb}
\end{eqnarray}
and can be found in the Appendix.
When we impose that such parameters agree with those of GR for any value
of $A$, we obtain two sets of compatible models, according to
their behavior in flat space-time, both with $\omega=0$:
\begin{itemize}
\item Gauge non-invariant models. These models have $\tau\neq 0$
and the corresponding term $\nabla_\mu A_\nu \nabla^\mu A^\nu$
breaks the $U(1)$ gauge invariance  in Minkows\-ki space-time. The
three possibilities we obtain in this case are:
$i)\;\sigma=-3\tau=-6\varepsilon$, $ii)\;\sigma=-2\tau=-2\varepsilon$, $iii)\;\sigma=\tau$.

\item Gauge invariant models. In this case we have $\tau=0$ and
the only term remaining in Minkowski space-time is the gauge
invariant one. The possibilities in this case are
$\sigma=m\varepsilon$ with $m=0,-2,-4$.
\end{itemize}
Notice that except for the $\sigma=\tau=0$ case (which is nothing
but GR plus Maxwell electromagnetism), all the considered cases
break gauge invariance in general space-times. Therefore, there
are six different classes of models which are indistinguishable
from GR by means of Solar System experiments and, therefore, do
not spoil the current bounds on the PPN parameters. To our knowledge best,
none of these models (apart from Maxwell's theory) had been considered
previously in the literature.

We will also use the present constraints on the variation
of the Newton's constant  which are given by  
$\dot G/G\;\lsim 10^{-13}$ yr$^{-1}$ \cite{Will}.

\section{Classical and quantum stability}

To study the existence of unstable classical modes and ghosts we
shall perform perturbations around a Minkowski background.
In addition, we will also
consider perturbations around the constant background vector field
 introduced in (\ref{vb}). This is possible because Minkowski
space-time is an admissible solution of the theory  in the case in
which the vector field takes a constant value, as that introduced
above. For this constant background, the vector field breaks
Lorentz invariance and we have a preferred frame defined as that
in which the vector field has only temporal component.
In this background we decompose the
perturbations in Fourier modes and solve the equations for the
corresponding amplitudes. That way, we obtain the dispersion
relation, which provides us with the propagation speed of the
modes, that is required to be real in order not to have
exponentially growing perturbations. As we explained above, here
we shall not care about superluminal propagation of the modes,
although it could be easily imposed at some point. Notice that
although we are assuming constant background, in practice, this
background could evolve on cosmological timescales. The effects of
such evolution could have important cosmological consequences
\cite{BM,BMEM}.

On the other hand, we define the energy for the modes as
\cite{Eling}:
\begin{equation}
\rho=\left<T_{00}^{(2)}-\frac{1}{8\pi
G}G_{00}^{(2)}\right>\label{defe}
\end{equation}
where $T_{\mu\nu}^{(2)}$ and $G_{\mu\nu}^{(2)}$ are the
energy-momentum tensor of the vector field and the Einstein's
tensor calculated up to quadratic terms in the perturbations and
$\left<\cdots\right>$ denotes an average over spatial regions.
Then, we insert the solutions obtained for the perturbations into
this expression and study under which conditions they are
positive.

Notice also that the preferred frame respects the invariance under
spatial rotations and therefore, in order to simplify the analysis
we can perform the usual split of the perturbations into spin-0
(scalar), spin-1 (vector) and spin-2 (tensor). For
simplicity, the longitudinal gauge has been chosen in
the calculations below, although we have checked that the final
results for the mode frequencies and energy densities
defined in (\ref{defe}) do not depend on the  gauge choice.

The scalar
perturbations of the vector field can be written as
$S_\mu=(S_0,\vec\nabla S)$ and the perturbed metric in the
longitudinal gauge is:
\begin{eqnarray}
ds^2=(1+2\phi)dt^2-(1-2\psi)\delta_{ij}dx^idx^j
\end{eqnarray}
For vector perturbations we have $V_\mu=(0,\vec{v})$ and the
metric is as follows:
\begin{equation}
ds^2=dt^2+2\vec{F}\cdot d\vec{x}dt-\delta_{ij}dx^idx^j
\end{equation}
with $\nabla\cdot\vec{v}=\nabla\cdot\vec{F}=0$. Although the
vector field does not generate tensor perturbations, we still can
have effects by the presence of the background vector field as we
shall see later. Let us first consider the scalar and vector
perturbations.

\vspace{0.5cm}

\subsection{Gauge non-invariant models}
\subsubsection{ Model I: $\sigma=-3\tau=-6\varepsilon$}

In this model, the Fourier components of the gravitational
potentials relate to those of the vector field as
follows\footnote{Hereafter we will measure the field in units of
$\sqrt{4\pi G}$.}:
\begin{eqnarray}
\phi_k&=&\frac{4\varepsilon A}{1-8\varepsilon A^2}\left[3\dot{S}_k-S_{0k}\right],\\
\psi_k&=&-\frac{4\varepsilon A}{1-8\varepsilon
A^2}\left[(1-32\varepsilon A^2)\dot{S}_k+S_{0k}\right],
\end{eqnarray}
whereas $S_{0k}$ and $S_k$ are related to each other by means of:
\begin{eqnarray}
S_{0k}&=&\frac{1}{k^2(1+8\varepsilon A^2)}\left[48\varepsilon
A^2k^2\frac{dS_k}{dt}\right.\nonumber \\
&-&\left.(1-8\varepsilon A^2)(1-16\varepsilon A^2)(1-64\varepsilon
A^2)\frac{d^3S_k}{dt^3}\right].
\end{eqnarray}
Thus, the problem is solved once we know the solution of $S_{k}$,
which happens to satisfy the following fourth order equation:
\begin{eqnarray}
\frac{d^4S_k}{dt^4}+\frac{k^2}{(1-16\varepsilon
A^2)(1-64\varepsilon
A^2)}\left[2(1-40\varepsilon
A^2)\frac{d^2S_k}{dt^2}+k^2S_k\right]=0.
\end{eqnarray}
This equation yields two independent modes:
\begin{equation}
S_k=C_1e^{-i\omega_1t}+C_2e^{-i\omega_2t},
\end{equation}
 with their respective
dispersion relations:
\begin{eqnarray}
\omega_1^2&=&\frac{k^2}{1-64\varepsilon A^2},\\
\omega_2^2&=&\frac{k^2}{1-16\varepsilon A^2}.
\end{eqnarray}
Then, in order to have stable solutions we need to satisfy the
following condition:
\begin{equation}
64\varepsilon A^2<1\label{NIGIIsm}
\end{equation}
because, in that case, both modes have real propagation speeds.

On the other hand, the energy density evaluated over the solutions
for each mode is given by:
\begin{eqnarray}
\rho_{\omega_1}^{(s)}&=&-384\varepsilon^2
A^2k^4\frac{(1-128\varepsilon
A^2)}{(1-64\varepsilon A^2)^2}\left|C_1\right|^2\\
\rho_{\omega_2}^{(1)}&=&384\varepsilon^2
A^2k^4\frac{(1-32\varepsilon A^2)}{(1+8\varepsilon
A^2)^2}\left|C_2\right|^2
\end{eqnarray}
These energies are both positive under the following constraint:
\begin{equation}
1<128\varepsilon A^2<4\label{NIGIIpe}
\end{equation}
Now, by combining (\ref{NIGIIsm}) and (\ref{NIGIIpe}) we find the
following viability condition for the scalar modes:
\begin{equation}
1<128\varepsilon A^2<2
\end{equation}

Concerning vector perturbations, we obtain the following relation
between the amplitudes:
\begin{equation}
\vec{F}_k=-\frac{16\varepsilon A^2}{1-16\varepsilon A^2}\vec{v}_k
\end{equation}
and both evolve as plane waves with the following dispersion
relation:
\begin{equation}
\omega^2=\frac{k^2}{1-16\varepsilon A^2}
\end{equation}
which leads to the stability condition:
\begin{equation}
16\varepsilon A^2<1
\end{equation}
Finally, the energy density associated to the vector perturbations
is given by:
\begin{equation}
\rho^{(v)}=8\varepsilon\frac{1-32\varepsilon
A^2}{(1-16\varepsilon)^2}k^2\left|\vec{v}_{0k}\right|^2
\end{equation}
From this expression we see that $\varepsilon$ must be positive
and we have to satisfy:
\begin{equation}
32\varepsilon<1
\end{equation}
in order not to have vector modes with negative energy.

In this model, according to the definition in the Appendix, $G=1$ 
and the constraints on the variation of $G$ do not set any limit
on the possible variation of $A$.

\subsubsection{ Model II: $\sigma=-2\tau=-2\varepsilon$}

\begin{figure}
\vspace{0.3cm}
\begin{center}{\epsfxsize=8.0 cm\epsfbox{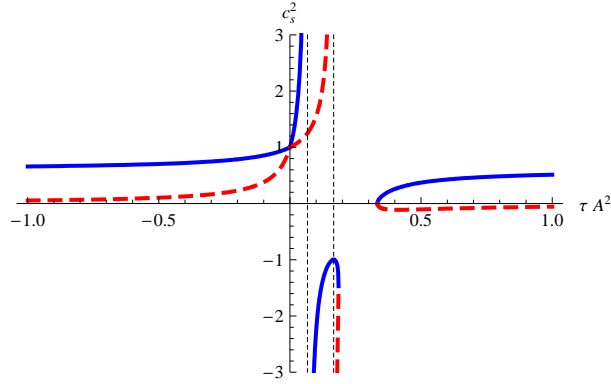}}
\caption{\small This plot shows the dependence of both modes
propagation speeds on $\tau A^2$ for the model II: blue for
$\omega_+$ and dashed-red for $\omega_-$. }\label{csmII}
\end{center}
\end{figure}

In this case, $\phi$ and $\psi$ are given in terms of the
perturbation of the vector field as follows:
\begin{eqnarray}
\phi_k&=&\frac{3\tau A}{1-3\tau
A^2}\dot{S}_k,\\
\psi_k &=&-3\tau A\frac{1-6\tau A}{1-3\tau A^2}\dot{S}_k.
\end{eqnarray}
On the other hand, the perturbation $S_{0k}$ can be expressed as:
\begin{eqnarray}
S_{0k}=&-&\frac{1}{2k^2(1-3\tau A^2)}\left[(1-6\tau A^2)(1-15\tau A^2)\frac{d^3S_k}{dt^3}\right.\nonumber\\
&-&\left.k^2(1+3\tau A^2)\frac{d S_k}{dt}\right].
\end{eqnarray}
Then, all the perturbations are given in terms of $S_k$, for which
we can obtain the following fourth-order differential equation:
\begin{eqnarray}
\frac{d^4S_k}{dt^4}+2k^2\frac{1-3\tau A^2}{(1-6\tau
A^2)(1-15\tau A^2)}\left[2(1-9\tau
A^2)\frac{d^2S_k}{dt^2}+k^2S_k\right]=0.
\end{eqnarray}
The solution of this equation is a superposition of two
independent modes:
\begin{equation}
S_k=C_+e^{-i\omega_+t}+C_-e^{-i\omega_-t},
\end{equation}
with their respective dispersion relations:
\begin{eqnarray}
\omega_\pm^2&=&\frac{k^2}{(1-6\tau A^2)(1-15\tau A^2)}\left[1-12\tau A^2+27\tau^2A^4\right.\nonumber\\
&&\left.\pm 3|\tau| A^2\sqrt{(1-3\tau A^2)(5-27\tau A^2)}\right].
\end{eqnarray}
Then, we have two modes with two different speeds of propagation
which depend on $\tau A^2$ as it is shown in Fig. \ref{csmII}. The
$\omega_-$-mode has real propagation speed for $\tau
A^2<\frac{1}{6}$, whereas for the $\omega_+$-mode to have real
propagation speed we need to satisfy either $\tau
A^2<\frac{1}{15}$ or $\tau A^2>\frac{1}{3}$. Therefore, the
necessary condition in order not to have instabilities is $\tau
A^2<\frac{1}{15}$. Finally, the degeneracy disappears for $\tau
A^2=0$ when both propagation speeds are 1, recovering thus the
usual result.

\begin{figure}
\vspace{0.3cm}
\begin{center}{\epsfxsize=8.0 cm\epsfbox{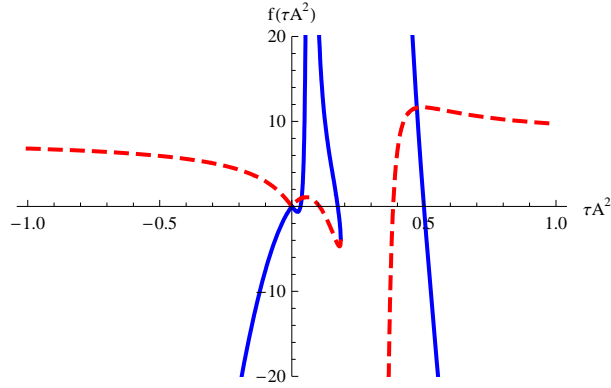}}
\caption{\small This plot shows the functions $f_\pm(\tau A^2)$
which determine the sign of the energy density for the scalar
modes in the model II: blue for $f_+$ and dashed-red for $f_-$
}\label{rhosmII}
\end{center}
\end{figure}

The energy density corresponding to each mode can be expressed as:
\begin{equation}
\rho^{(s)}_\pm=\tau f_\pm(\tau A^2)k^4\left|C_\pm\right|^2
\end{equation}
where $f_\pm(\tau A^2)$ are the functions plotted in Fig.
\ref{rhosmII}. Notice that $f$ and $\tau$ must have the same sign
for the energy density to be positive. We find the following
condition in order to have positive energy density for both modes:
\begin{equation}
\tau A^2\in(0.033,0.105)\cup(0.383,0.5)
\end{equation}

For this model, the vector perturbation on the metric relates to
that of the vector field by means of:
\begin{equation}
\vec{F}_k=-\frac{6\tau A}{1-6\tau A^2}\vec{v}_k
\end{equation}
which evolve as:
\begin{eqnarray}
\vec{v}_k=\vec{v}_{0k}e^{-i\omega_vt}
\end{eqnarray}
where $\vec{v}_{0k}\cdot \vec k=0$ and 
\begin{equation}
\omega_v^2=\frac{1-3\tau A^2}{1-6\tau A^2}k^2
\end{equation}
From this expression we obtain that for $\tau A^2<\frac{1}{6}$ the
propagation speed is real.

The energy density corresponding to the vector perturbations is:
\begin{eqnarray}
\rho^{(v)}=6\tau k^2\frac{1-12\tau A^2+18\tau^2A^4}{(1-6\tau
A^2)^2}\left|\vec{v}_{0k}\right|^2
\end{eqnarray}
One can easily verify that this expression is positive if:
\begin{equation}
\tau A^2\in \left[0,c_-\right]\cup\left[c_+,\infty\right)
\end{equation}
with $c_\pm=\frac{1}{3}(1\pm\frac{1}{\sqrt{2}})$. Note that
$\tau>0$ is a necessary condition and that the singular value
$\tau A^2=\frac{1}{6}$ is not contained in the interval.

In this case, $G=1+6\tau A^2$. The present constraints on the variation
of the Newton's constant  will translate into 
a limit on the possible variation of $A$ which will depend on the 
present cosmological value of $A$.

\subsubsection{ Model III: $\sigma=\tau$}

The perturbations of this model propagate at the speed of light so
there are no classically unstable modes. However, the scalar perturbations are
not just plane waves, but they have also a growing mode:
\begin{eqnarray}
S_{0k}&=&\left[-ik\lambda(D_0+ikD)t+D_0\right]e^{-ikt}\nonumber\\
S_{k}&=&\left[\lambda(D_0+ikD)t+D\right]e^{-ikt}\label{Gsol}
\end{eqnarray}
with
\begin{equation}
\lambda=\frac{\varepsilon+2\tau(2\varepsilon+\tau)A^2}{\varepsilon+\tau+2\tau(2\varepsilon+\tau)A^2}\nonumber
\end{equation}

In principle,  this solution can give rise to both positive or
negative energies depending on the value of the amplitudes.
However, there exists a way to get this difficulty around
which is motivated by the fact that the action corresponding
to this model can be rewritten as that of Maxwell's electromagnetism
in the Gupta-Bleuler formalism, that is:
\begin{eqnarray}
S&=&\int d^4x\sqrt{-g}\left[-\frac{1}{16\pi G}R-\frac{1}{4}F_{\mu
\nu}F^{\mu\nu}+\frac{\xi}{2}\left(\nabla_\mu
A^\mu\right)^2\right]
\label{GB}
\end{eqnarray}

Notice that this action is the physically relevant action
in the covariant formalism, since the energy of the modes and
all the observables are calculated from (\ref{GB})
(see for instance \cite{Itzkynson}). In order to get rid of
the negative energy modes, it is necessary to restrict the 
Hilbert space of the theory, by imposing the (Lorentz) condition 
$\langle \phi\vert\partial_\mu A^\mu\vert\phi\rangle=0$, 
which determines the physical states $\vert \phi\rangle$.
As shown in \cite{BMEM}, this is equivalent to remove the 
growing mode i.e.  $D_0=-ikD$ 
 so that perturbations of the vector field propagate
as pure plane waves. Then, using this condition we obtain that, as
in the electromagnetic case, in the restricted Hilbert space,
the energy of the scalar modes is identically zero.
 A detailed treatment of the quantization
for this model is performed in \cite{BMEM}.

The solution of the vector perturbation of the vector field is:
\begin{equation}
\vec{v}_k=\vec{v}_{0k}e^{-ikt}
\end{equation}
 Moreover, the
vector perturbation of the metric vanishes.

The energy density associated to the vector perturbations is:
\begin{equation}
\rho^{(v)}=(2\varepsilon+\tau)k^2\left|\vec{v}_{0k}\right|^2
\end{equation}
which is positive if $2\varepsilon+\tau>0$.

In this model we have again $G=1$ 
and no constraints on the variation of $A$ can be established.

\vspace{0.5cm}

\subsection{ Gauge invariant models: $\tau=0$, $\sigma=m\varepsilon$,
$m=0,-2,-4$}

As the case $m=0$ is nothing but Einstein's gravity plus
electromagnetism (which can be easily seen to satisfy all the viability
conditions), we shall focus just
on the cases $m=-2,-4$. In such cases, we can obtain the following
relations between the perturbations:
\begin{eqnarray}
\dot{S}_k&=&\frac{1+(m+3)m\varepsilon A^2}{m\varepsilon
A(1+2m\varepsilon A^2)}\psi_k,\nonumber\\
S_{0k}&=&\frac{1-m\varepsilon A^2(3-2m^2\varepsilon
A^2)}{m\varepsilon A(1+2m\varepsilon A^2)}\psi_k,\nonumber\\
\phi_k&=&\psi_k-2m\varepsilon A\dot{S}_k.\label{GIamp}
\end{eqnarray}
Therefore, all the perturbations can be immediately obtained once
we know the solution of $\psi_k$ which happens to evolve as:
\begin{equation}
\psi_k=C_ke^{-i\omega_s t},
\end{equation}
with:
\begin{equation}
\omega_s^2=\frac{1}{3}\frac{3+2(4+m)m\varepsilon
A^2}{1+2m\varepsilon A^2}k^2.
\end{equation}
Then, the classical stability of these modes is guaranteed in the
range:
\begin{equation}
\varepsilon
A^2\in\left(-\infty,-\frac{1}{2m}\right)\cup\left(-\frac{3}{2(4+m)m},\infty\right).\nonumber
\end{equation}
Notice that the second interval vanishes for the model with
$m=-4$.

When we insert the corresponding solutions into the energy density
we obtain:
\begin{equation}
\rho^{(s)}=\frac{3+4m\varepsilon A^2\left(3+(4+m)m\varepsilon
A^2\right)}{2\pi G(1+2m\varepsilon A^2)}\;k^2\left|C_k\right|^2.
\end{equation}
This energy density is positive for:
\begin{eqnarray}
\varepsilon A^2\in(-\infty,a_+)\cup\left(-\frac{1}{2m},a_-\right),
\end{eqnarray}
with:
\begin{equation}
a_\pm=-\frac{1}{2m\left(1\pm\sqrt{-\frac{1+m}{3}}\right)}.
\end{equation}
Notice that $a_-$ is $\infty$ for the model $m=-4$.

On the other hand, the vector perturbations of the field evolve as
plane waves
$\vec{v}_k=\vec{v}_{0k}e^{-i\omega_vt}$ with the
following dispersion relation:
\begin{equation}
\omega_v^2=\frac{1+(2+\frac{1}{2}m)m\varepsilon A^2}{1+2
m\varepsilon A^2}k^2,
\end{equation}
which imposes the condition:
\begin{equation}
\varepsilon
A^2\in\left(-\infty,-\frac{1}{2m}\right)\cup\left(-\frac{1}{(2+\frac{1}{2}m)m},\infty\right)\nonumber
\end{equation}
in order to have real propagation speed. Notice that for 
$m=-4$ the second interval vanishes.

Besides, the vector perturbation of the metric relates to
$\vec{v}$ by means of:
\begin{equation}
\vec{F}_k=\frac{2m\varepsilon A}{1+2m\varepsilon A^2}\vec{v}_k.
\end{equation}
These solutions for the vector perturbations lead to the following
expression for the energy density:
\begin{eqnarray}
\rho^{(v)}=\varepsilon\frac{1+4m\varepsilon
A^2\left(1+(1+\frac{1}{4}m)m\varepsilon A^2\right)}{(1+2
m\varepsilon A^2)}\;k^2\left|\vec{v}_{0k}\right|^2
\end{eqnarray}
The requirement for this energy to be positive is:
\begin{equation}
\varepsilon A^2\in(0,b_+)\cup\left(-\frac{1}{2m},b_-\right),
\end{equation}
where
\begin{equation}
b_\pm=-\frac{1}{m(2\pm\sqrt{-m})}
\end{equation}
Note also that for $m=-4$, $b_-$ becomes $+\infty$.

In this case we get: $G=1-\varepsilon m(4+m)A^2$, so the only model
with $G\neq 1$ is that with $m=-2$. 

\section{Gravitational Waves}

At first glance, one may think that as the vector field does not
generate tensor modes at first order,  gravitational waves will
not be affected. However, the presence of a constant value of the
vector field in the background can modify the speed of propagation
of  tensor perturbations. For the general vector-tensor action
(\ref{Action}) we have the following dispersion relation:
\begin{equation}
\omega_t^2=\frac{k^2}{\sqrt{1+2(\sigma-\tau)A^2}}
\end{equation}
for both $\oplus$ and $\otimes$ polarizations. Therefore, the
speed of  gravitational waves is modified in the presence of the
vector field,  recovering the usual value for $A=0$. Thus, if
$2(\sigma-\tau)A^2>-1$ we do not have unstable modes. In
particular, the constraint $\sigma-\tau\geq 0$ is a sufficient
condition (although not necessary) which is independent of the
background vector field.

On the other hand, the energy density associated to the tensor
perturbations is also modified by the presence of the background
vector field:
\begin{eqnarray}
\rho^{(t)}=\frac{k^2}{32\pi
G}\frac{1+4(\sigma-\tau)A^2}{1+2(\sigma-\tau)A^2}
\left(\vert C_\oplus\vert^2+\vert C_\otimes\vert^2\right)
\end{eqnarray}
where $C_\oplus$, $C_\otimes$ are the amplitudes of the
corresponding graviton polarizations.

Then, in order not to have modes with negative energy density we
need either $2(\sigma-\tau)A^2<-1$ or $4(\sigma-\tau)A^2>-1$.
These conditions combined with the classical stability condition
lead to the constraint $4(\sigma-\tau)A^2>-1$. For models I and II
this condition reads $2\varepsilon A^2<1$ and $12\tau A^2<1$
respectively. On the other hand, Model III has $\sigma=\tau$ so
gravitational waves are unaffected. Finally, for the gauge
invariant models we obtain $\varepsilon A^2<\frac{1}{4m}$.

\begin{table*}
{\scriptsize \hspace{0.3cm}\begin{tabular}{|c|c|c|c|c|}
\hline
  & & & &\\
 & Model I  & Model II  & Model III & Gauge invariant models \\

  & & & &\\
 &   &   &  & $m=-2,\, -4$ \\

& & & &\\
\hline & & & & \\Classical Stability & $64\varepsilon A^2<1$ &
 $15\tau A^2<1$ &Always & $
\varepsilon
A^2\notin\left[-\frac{1}{2m},-\frac{1}{\left(2+\frac{1}{2}m\right)m}\right]
$

\\& & & & \\
\hline & & & &\\ Gravitational Waves & $2\varepsilon A^2<1$ &
 $12\tau A^2<1$ & --- & $\varepsilon A^2<\frac{1}{4m}$
\\

& & && \\
\hline & & & &\\ Quantum Stability & $1<128\varepsilon A^2<4$ &
 Not viable & $\partial_\mu A^\mu=0$ and $2\varepsilon+\tau >0$ & $\varepsilon A^2\in(0,b_+)
\cup\left(-\frac{1}{2m},b_-\right)$\\
& & & & \\
\hline  & & & &\\Viability condition & $1<128\varepsilon A^2<2$ &
Not viable &$\partial_\mu A^\mu=0$ and $2\varepsilon+\tau >0$ &
Incompatible

\\ & & & &\\
\hline
\end{tabular}\vspace{0.5 cm}}
\footnotesize{{\bf Table 1:} In this table we summarize the conditions
obtained in order to have both classical and quantum stability for
the models with the same set of PPN parameters as GR studied in
this work. The $m=0$ gauge invariant model satisfies all the
viability conditions.}
\end{table*}

\section{Conclusions}
We have studied the viability conditions for unconstrained
vector-tensor theories
of gravity. In Table I we summarize the different
conditions obtained for the different models analyzed.\footnote{Notice 
that we have only considered models with exactly the 
same PPN parameters as GR, if we relax this condition and 
we only demand compatibility with current experimental 
limits on PPN parameters then the range of viable models could 
be much larger.}  
If we concentrate only on classical
stability and local gravity constraints, we find that there are
 different types of vector-tensor theories which can be made viable
for certain values of the coefficients. Thus for example: Model I with
$\epsilon<0$ or Model II with $\tau<0$ for arbitrary values of $A$.

 However if we also impose that the models are
free from ghosts, we have found that only theories of the Maxwell
type or Maxwell plus a gauge-fixing term can be made compatible
with all the consistency conditions for arbitrary $A$. Notice that
on small (sub-Hubble) scales, such vector-tensor theories are
indistinguishable from General Relativity, however it has been
shown (see \cite{BM,BMEM} for different examples) that in a
cosmological scenario, the presence of the (dynamical) background
vector field could have important consequences. 
Indeed in \cite{BMEM} it has been shown that the electromagnetic theory
with a gauge fixing term can explain in a natural way  the 
existence and the smallness of the
cosmological constant, solving in this way the problem of 
establising what is the fundamental nature of dark energy.
Present and
forthcoming astrophysical and cosmological observations could thus
help us determining what is the true theory for the gravitational
interaction.

\vspace{0.3cm}

{\em Acknowledgments:}
This work has been  supported by
DGICYT (Spain) project numbers FPA 2004-02602 and FPA
2005-02327, UCM-Santander PR34/07-15875, CAM/UCM 910309 and
MEC grant BES-2006-12059.

\section*{Appendix: PPN parameters for  vector-tensor theories of gravity.}

The PPN parameters for the vector-tensor theory (\ref{Action}) are
given by \cite{Will}:
\begin{eqnarray}
\gamma&=&\frac{1+4\omega
A^2\left(1+2\frac{2\omega+\sigma-\tau}{2\varepsilon+\tau}\right)}{1-4\omega
A^2\left(1-\frac{8\omega}{2\varepsilon+\tau}\right)}\\
\beta&=&\frac{1}{4}(3+\gamma)+\frac{1}{2}\Theta\left[1+\frac{\gamma(\gamma-2)}{G}\right]\nonumber\\
\alpha_1&=&4(1-\gamma)\left[1+(2\varepsilon+\tau)\Delta\right]+16\omega
A^2\Delta a\nonumber\\
\alpha_2&=&3(1-\gamma)\left[1+\frac{2}{3}(2\varepsilon+\tau)\Delta\right]+8\omega
A^2\Delta a-2\frac{bA^2}{G}\nonumber\\
\alpha_3&=&\zeta_1=\zeta_2=\zeta_3=\zeta_4=0\nonumber
\end{eqnarray}
with:
\begin{eqnarray}
&\Theta&=\frac{(1-4\omega
A^2)(2\varepsilon+\sigma-2\omega)}{(1-4\omega
A^2)(2\varepsilon+\tau)+32\omega^2A^2}\nonumber\\
&\Delta&=\frac{1}{2A^2(\sigma-\tau)^2-(2\varepsilon+\tau)
\left[1-4A^2(\omega+\sigma-\tau)\right]}\nonumber\\
\nonumber \\
&a&=(2\varepsilon+\tau)(1-3\gamma)+2(\sigma-\tau)(1-2\gamma)\nonumber\\
\nonumber \\
&b&=\left\{\begin{array}{c}
(2\omega+\sigma-\tau)\left[(2\gamma-1)(\gamma+1)+\Theta(\gamma-2)\right]\\
-(2\gamma-1)^2(2\omega+\sigma)\left(1-\frac{2\omega+\sigma}{\tau}\right)\;\;\;\;\;\;\;\;\;\;\;\tau\neq0\\\\
\;\;\;\;\;\;\;\;\;\;\;\;\;\;0\;\;\;\;\;\;\;\;\;\;\;\;\;\;\;\;\;\;\;\;\;\;\;\;\;\;\;\;\;\;\;\;\;\;\;\;\;\;\;\;\;\;\;\;\;\tau=0
\end{array} \right.\nonumber
\end{eqnarray}
Moreover, it is possible to define an effective Newton's constant given by:
\begin{eqnarray}
G_{eff}&\equiv& G\left[\frac{1}{2}(\gamma+1)+6\omega
A^2(\gamma-1)\right.\nonumber \\
&-&\left.2A^2(\sigma-\tau)(1+\Theta)\right]^{-1}.
\end{eqnarray}
The above expressions are obtained assuming $G_{eff}=1$.

\end{document}